%% file: main.tex
\documentclass[conference]{IEEEtran}

\usepackage{balance}

\IEEEoverridecommandlockouts 
\usepackage[pdftex,dvips]{graphicx}
\usepackage{algorithmic}
\usepackage{algorithm}
\usepackage{listings}
\usepackage[OT4,T1]{fontenc}
\usepackage[cmex10]{amsmath}
\usepackage{url}
\usepackage{multirow}

\title{Incorporating Text Analysis into Evolution of Social Groups in Blogosphere}

\author{\IEEEauthorblockN{
Bogdan Gliwa, Anna Zygmunt, Stanis{\l}aw Podg{\'o}rski}
\IEEEauthorblockA{Department of Computer Science\\
AGH University of Science and Technology\\
30-059 Krak\'ow, Poland\\
\{bgliwa,azygmunt\}@agh.edu.pl, pstanisl@student.agh.edu.pl}
}

\begin{document}
\maketitle              

\begin{abstract}
Data reflecting social and business relations has often form of network of connections between entities (called social network). In such network important and influential users can be identified as well as groups of strongly connected users. Finding such groups and observing their evolution becomes an increasingly important research problem. One of the significant problems is to develop method incorporating not only information about connections between entities but also information obtained from text written by the users. Method presented in this paper combine social network analysis and text mining in order to understand groups evolution.
\end{abstract}

\input{Introduction}

\input{RelatedWork}

\input{Model}
\input{Results}

\input{Conclusion}

\bibliographystyle{IEEEtran}
\bibliography{biblio,blogs,groups,snajkaz}

\end{document}

%% file: Introduction.tex
\section{Introduction}

Nowadays, elements of our everyday life move increasingly to the virtual reality: we write blogs or comment on someone else's posts, participate in discussions on  forums, we exchange our opinions on fanpages of telecommunications companies and banks whose products we use. Everywhere we leave traces of our activity, which can be analyzed and ably combined with each other. Trading companies and banks might be interested in finding active or influential people in their environment and to offer them a new product in hope that it will be proposed by them to many others. Identification on time disgruntled people on banks or telecommunications companies fanpages will allow to respond quickly and prevent the spread of discontent.

The data about different types of dependencies can be modeled as a network of relationships and its structure can be analyzed using Social Network Analysis methods (e.g. for finding important nodes). Such a network, however, is not homogeneous and one can distinguish groups of people, for example, who more often exchange opinions. Such groups frequently are formed around important individuals and, for various reasons, the groups continue to exist or not, grow, shrink or can be joined with other groups. To understand causes of such events, which significantly affect the behavior of groups, it is important to include information that we can extract from the content of opinions or comments that are left behind. If the group talk about the same topics, does it affect its longer duration? Or, perhaps has a variety of discussed topics stronger impact on the duration of groups? How the themes of discussion are changing in a group? Is a small group with a strong leader more durable than a large one with few strong individuals?

Such knowledge derived from open sources can be combined, for example, with information about the history of bank transfers or loans as well as data about phone calls. Methods and algorithms proposed in the paper have been tested on one of the largest and highly dynamic polish blogosphere: salon24.pl, in which the main topic of discussion are political issues. However, they can be applied to other social media such as, for example, Twitter.

%% file: RelatedWork.tex
\section{Related work}

\subsection{Blogosphere and Social network analysis}

Internet social media (e.g.  blogs, forums, media sharing systems, microblogging, social networking, wikis) has revolutionized the Internet and the way of communication between people. Among them, blogs play a special role in creating opinions and information propagation. Author gives opinions on some themes or describes interesting events and readers comment on these posts. Posts can be categorized by tags. A very important element of blogs is the possibility of adding comments, which allows discussions. 
Basic interactions between bloggers are writing comments in relation to posts or other comments. The relationships between bloggers are very dynamic and temporal: lifetime of a post is very short \cite{AgarwalLiu:09}.

Based on blogs, posts and comments, we can build network, which can be analysed by Social Network Analysis (SNA) methods \cite{carrington2005}. The SNA approach provides measures (SNA centrality measures) which make it possible to determine the most important or influential nodes (bloggers) in the network. Around such bloggers, the groups are forming, sharing similar interests.

\subsection{Groups in social networks}

Groups (or communities) are sets of nodes that are relatively densely connected to each other but sparsely connected to other nodes in network \cite{Fortunato2010}. Many methods of finding groups exist in literature - one of the most popular ones is the CPM method (Clique Percolation Method) \cite{palla2005}, which allows to extract overlapping groups i.e. groups that can have shared nodes with other groups. 

Considering the dynamic nature of various social media, a growing interest in developing algorithms for extracting communities that take into account the dynamic aspect of the network has been observed.

A method of tracking groups over time was proposed in \cite{greene10tracking}. First, a division into time steps is carried out. At each step, the graph is created and groups are extracted. Groups from consecutive time steps are matched using the Jaccard index (value of this measure above predefined threshold means a continuation for analysed group). Palla et al. in \cite{Palla07} identified basic events that may occur in the life cycle of the group: growth, merging, birth, construction, splitting and death. 

For further analysis, different characteristics, describing the communities and their transformation in time \cite{xu2004}, are calculated, which concerns the comparison of the strength of internal relations of group members with their external connections with nodes outside the group, density of connections in the group or stability of the membership in time.

\subsection{Methods of text mining}

Text classification is one of major goals of Text Mining  \cite{generalTextMining}. It involves extracting similar documents, inferring text topic and searching documents based on topic criteria. 

Most text mining methods focus on text preprocessing (eg. stop words removal, words stemming and lemmatization) and converting input into structural representation \cite{textAnalyticsSocial}. Each word is represented as a separate entity assigned with a weight of the word importance, and thus it also allows to easily extract keywords. Algorithm TF-IDF (Term Frequency - Inverted Document Frequency)  is one of the most popular weighting method  \cite{textAnalyticsSocial}.
It is based on the assumption that the importance of a word is proportional to number of occurrences of this word in a document, and inversely proportional to number of documents in which the word occurred. However, using only keywords to classify texts fails to find connection between semantically convergent documents that utilize different vocabularies, and more complex methods need to be applied such as Topic Modeling \cite{miningTextDimensionality}.

Topic Modeling \cite{lsi} is a statistical technique that uncovers abstract "topics" that occur in a collection of documents. "Topic" is a set of words that tend to co-occur in multiple documents, and, therefore, they are assumed to have similar semantics. Main benefit of this model is that instead of using words from pattern to search for similar documents, words from topic are used, and therefore similar texts can be discovered even if they use different vocabulary.


Entirely different approach to uncovering documents semantics involves human input and it is called tagging \cite{tag}. Tags can be assigned either by author or by community in a process called crowdsourcing \cite{crowdsourcing}. Number of tags assigned to a document may be large, and, therefore, it is imperative that a proper grouping and selection mechanism is implemented. 

\subsection{Text mining in the context of social network analysis}

Existing research utilizing both SNA and Text Mining are mainly focused on very narrow cases. Aggarwal and Wang in \cite{analyticsNetworks} provided broad overview of text mining methods useful for social networks analysis.
Tuulos and Tirri in \cite{chats} analysed IRC (Internet Relay Chat) communication network to discover and verify chat channels topics.
Agrawal et al. in \cite{newsgroups} used text mining methods to split social group into protagonists and antagonists.
Caverlee and Webb in \cite{myspace} used automatic classification methods based on keywords extraction and Topic Modeling to confirm personal information provided by Myspace users.

%% file: Model.tex
\section{Analysis topics of groups and their impact on group behaviour}
In this section we provide the concept of methods used to further analysis. The social network from whole data range is divided into series of time slots and each time slot contains static snapshot of network from defined period of time. In every time slot we extract groups and then find their dynamics in time. Irrespectively, we also discover topics in texts of comments and posts. Afterwards, we try to match topics for groups based on topics of comments and posts written by members of groups between themselves. Next, we analyse relations between topics of groups and behaviour of groups.

\subsection{Groups in dynamic social network}
Groups in each time slot were detected  using the CPM \cite{palla2005} method (directed version of CPM from CFinder \footnote{http://www.cfinder.org/}).
Groups from neighbouring time slots can be matched in order to find continuation of groups from different time. For this purpose, the SGCI (Stable Group Changes Identification) \cite{GliwaSZBKK12} method was employed. The algorihm consists of four main steps: 
identification of  short-lived groups in each separated time slot; identification of  group continuation (using modified Jaccard measure), separation of the stable groups (lasting for a certain time interval) and the identification of types of group changes (transition between the states of the stable group). 
The SGCI method identifies following event types: 
\begin{itemize}

\item {\bf split}, occurs when group divides into several groups in next time slot,

\item {\bf deletion}, similar to split, but it happens when small group detaches from significantly bigger one,

\item {\bf merge}, when several groups in the previous time slot join together and create larger group,

\item {\bf addition}, similar to merge, but it takes place when small group attaches to significantly bigger group,

\item {\bf split\_merge}, when for the predecessor group the event is split and for the succesor group of given transition the event is merge in the same time,

\item {\bf decay}, the total disintegration of the group - the group does not exist in the next time slot,

\item {\bf constancy} means simple transition without significant change of the group size, 

\item {\bf change\_size} - simple transition with the change of the group size.

\end{itemize}

More detailed description of this method is provided in \cite{GliwaSZBKK12}.

\subsection{Finding topics of groups}
For texts of posts and comments we employed methods of text mining in order to discover topics. Topics were extracted using 3 different methods:
\begin{itemize}
\item TF-IDF keywords - words with the highest TF-IDF scores,
\item Topic Modeling - topics extracted with LDA algorithm,
\item Tags provided by post authors.
\end{itemize}

Keywords set for Topic Model is assumed to be a set of the most significant words for topics inferenced for messages.

We compared these methods between themselves using {\it similarity} measure:
\begin{equation*}
	 similarity(S_1,S_2) = \frac{|S_{1} \cup S_{2}|}{\min(|S_{1}|,|S_{2}|)}
\end{equation*}
where: 
$S$ -- keywords set,
$|S|$ -- number of elements in $S$.

For each group we can also assign set of topics discussed by its members. The topics are inferred based on posts and comments written by members of groups. We focused mostly on topic modelling as this method provides the highest level of abstraction from presented methods. Only topics that were present in more than 5\% messages for groups were taken into consideration.

We defined {\it topic exploitation} for given topic and group as a ratio between number of group messages on certain topic and all messages for this group:
\begin{equation*}
		 topicExploitation_{k} = \frac{|T_{k}|}{\sum\limits_{i=1}^n |T_{i}|}
\end{equation*}
where:
$T_{k}$ -- set of messages (posts and comments) for which topic with number $k$ was inferenced,
$n$ -- number of all topics,
$|T_{x}|$ -- amount of elements in $T_{x}$.

\subsection{Topics changes in groups} \label{topicChanges}
To describe topics changes during transition between groups, we introduced following metrics:
\begin{itemize}
\item {\it Change in topic exploitation} for $m$-th group after transition $t$  from time slot $n$ to $n+1$ is calculated as:
\begin{equation*}
	c_{m,n,t} = \sum\limits_{i} \sum\limits_{k} [g_{m,n,i} - g_{k,n+1,i}\cdot f(m,n,k,t)]
\end{equation*}
where:
$i$ is a number of topic,
$g_{m,n,i}$ is the topic exploitation of $i$-th Topic for $m$-th group in $n$-th time slot,
$f$ is function returning 1 if $k$-th group in slot $n+1$ is a continuation of $m$-th group from slot $n$ and this transition has event type $t$.

\item {\it Maximal positive change of single topic} (how much a topic gained) for $m$-th group after transition $t$ from time slot $n$ to $n+1$ is defined as:
\begin{equation*}
	mpc_{m,n,t} = \max\limits_{i} \sum\limits_{k} [\epsilon(g_{m,n,i} - g_{k,n+1,i}\cdot f(m,n,k,t))]
\end{equation*}
where:
$\epsilon$ is a function returning the argument when the argument is {\bf positive}, otherwise 0;
other symbols were explained for {\it Change in topic exploitation} measure.

\item {\it Maximal negative change of single topic} (how much a topic lost) for $m$-th group after transition from time slot $n$ to $n+1$ was calculated as:
\begin{equation*}
	mnc_{m,n,t} = \max\limits_{i} \sum\limits_{k} [\theta(g_{m,n,i} - g_{k,n+1,i}\cdot f(m,n,k,t))]
\end{equation*}
where:
$\theta$ is a function returning the argument when the argument is {\bf negative}, otherwise 0;
other symbols were explained for {\it Change in topic exploitation} measure.

\end{itemize}

Using above metrics we can analyse influence of different evolution types on topics change. Therefore, for each evolution type the average values of above defined measures for all groups are evaluated and we refer to them as {\it Average overall change in topic exploitation}, {\it Average maximal positive change of single topic} and {\it Average maximal negative change of single topic} respectively.

For above metrics, evolution events were taken into consideration only if there were at least 10 such events in selected time period.

\subsection{Migrations of users depending on topics}\label{topicDivergence}

To analyse difference in topics between given user and given group, we defined {\it topic divergence}, which has the following form:
\begin{equation*}
 m_{t} = t_{group} - t_{user} = \sum\limits_{i=1}^n{|(topic_{i,user} - topic_{i,group})|} 
\end{equation*}

where:
$n$ is a number of all topics in model (350),
$t_{group}$ is set of weights of each topic for given group,
$topic_{i,group}$ -- weight of $i$-th topic for given group,
$t_{user}$ is set of weights of each topic for given user,
$topic_{i,user}$ is weight of $i$-th topic for given user.

It's worth noting that minimal value of $m_{t}$ is 0.0 when user and a group has identical weight for every topic and maximal value is 2.0 when they are totally different.
Maximum value of 2.0 is connected with the fact that group might cover topic $X$ in 100\% and user might cover topic $Y$ in 100\%, and therefore difference between group and user on topic $X$ is 100\% and on topic $Y$ is also 100\% which adds up to 200\%.

Using this measure, we are trying to investigate relations between {\it topic divergence} and migrations of users (leaving and joining to groups). For this purpose the following measures are utilized:
\begin{itemize}
\item {\it Probability of leaving the group}. We assumed that potentially any member can leave the group. This value is calculated as:
	
\begin{equation*}
	P_{l}(m) = \frac{|leavers_{m} \cap candidates_{m}|}{|candidates_{m}|}
\end{equation*}
where:
$leavers_{m}$ are users that in fact left any group and had the value of {\it topic divergence} measure equals $m$;
$candidates_{m}$ are members of groups that have {\it topic divergence} = $m$.
		
\item {\it Probability of joining the group}. When considering topic measure we assumed that candidates for joining are all users that were active in previous time slot. This value is calculated as:
	
\begin{equation*}
	P_{j}(m) = \frac{|joiners_{m} \cap candidates_{m}|}{|candidates_{m}|}
\end{equation*}
where:
$joiners_{m}$ are users that in fact joined any group and had the value of {\it topic divergence} measure equals $m$;
$candidates_{m}$ -- users active in previous time slot with {\it topic divergence} = $m$.
\end{itemize}

While calculating joiners and leavers sets we considered all group continuations to be a single group. The reason for that is to prevent {\it deletion} event to distort results - if a group splits into multiple small groups and we are assuming that anyone from the group can leave, then we will get very high accuracy from each event when huge group changes into small group.

It is worth noting that only both values - probability and histogram with migrations can provide us with complete information.
Probability alone strongly depends on test case - if only 1 user had measure value = X and this user migrates then probability of migration for measure=X will be 100\%, even if 100 different users migrated but they all had measure value=Y just as rest 10000 users, and thus probability for measure value=Y will be 1\%.
Without histogram we could not tell if any of those cases are marginal.

Analogically histogram itself can tell us only for which value there are the most migrations.

%
%
%
%

%% file: Results.tex
\section{Description of experiments}

\subsection{Data set}

The analysed data about blogs was retrieved from the portal www.salon24.pl, which is dedicated especially to political discussions, but also subjects from other domains may be brought up. The data consists of 26 722 users (11~084 of them have their own blog), 285~532 posts and 4 173 457 comments within the period 1.01.2008 - 31.03.2012. Presented results were conducted on whole dataset - from 1.01.2008 to 31.03.2012.  The analyzed period was divided into time slots, each lasting 7~days and neighboring slots overlap each other by 1 days. In the examined period there are 259 time slots. In each slot we used the comments model, introduced in \cite{Gliwa:2012b} - the users are nodes and relations between them are built in the following way: from user who wrote the comment to the user who was commented on or if the user whose comment was commented on is not explicitly referenced in the comment (by using @ and name of author of comment) the target of the relation is the author of post.

\subsection{Number of groups}

The number of communities, with given size, for different value of {\it k} (parameter for CPM algorithm) is presented in table \ref{tab:groupSizes}. The {\it k} parameter determines the minimum group size (e.g. {\it k} equals 3 means that groups should consist of 3 or more members). The larger value of {\it k}, the smaller size of the biggest group.
As we can notice, small groups outnumber other ones for each {\it k}. Furthermore, for {\it k} equals 6 the quantity of groups is much lower than for other values of {\it k} parameter.

\begin{table}[ht]\normalsize
\centering
\begin{tabular}{cccc}
size & k=4 & k=5 & k=6 \\
\hline
< 5 & 1596 & 0 & 0 \\
5 -- 6 & 384 & 2372 & 0 \\
6 -- 7 & 207 & 632 & 584 \\
7 -- 8 & 113 & 255 & 149 \\
8 -- 9 & 88 & 139 & 86 \\
9 -- 10 & 50 & 63 & 39 \\
10 -- 50 & 289 & 332 & 199 \\
50 -- 100 & 25 & 54 & 30 \\
100 -- 200 & 59 & 96 & 6 \\
> 200 & 172 & 17 & 0 \\
\end{tabular}
\caption{Numbers of groups with defined size.}
\label{tab:groupSizes}
\vspace{-0.3cm}
\end{table}


\subsection{Evolution events}

Table \ref{tab:transitions} contains number of different evolution events in dataset for different values of {\it k}. We can observe for {\it k} equal 4 or 5 that the most popular events are {\it addition} and {\it deletion}, but for {\it k} equal 6, the most frequent events are {\it merge} and {\it split} (events similar to {\it addition} and {\it deletion}). The reason is that for {\it k} equal 4 or 5, there is a lot of small groups and there are also very huge groups (which not happens for {\it k} equal 6). In further analysis, we focus on groups extracted for parameter {\it k} equal 5 from the CPM method. 

\begin{table}[ht]\normalsize
\centering
\begin{tabular}{l*{6}{c}r}
type & k=4 & k=5 & k=6 \\
\hline
change\_size & 699 & 470 & 195 \\
constancy & 257 & 100 & 42 \\
merge & 428 & 439 & 434 \\
split & 323 & 409 & 397 \\
addition & 1091 & 2070 & 197 \\
deletion & 1115 & 2040 & 188 \\
\end{tabular}
\caption{Numbers of evolution events.}
\label{tab:transitions}
\vspace{-0.3cm}
\end{table}


\subsection{Convergence of different message topic extraction methods}

This experiment covered comparison of different messages topic inference methods: TF-IDF keywords, topic modelling and tags provided by users.

\begin{figure}[ht]
	\centering
	\includegraphics[width=\linewidth]{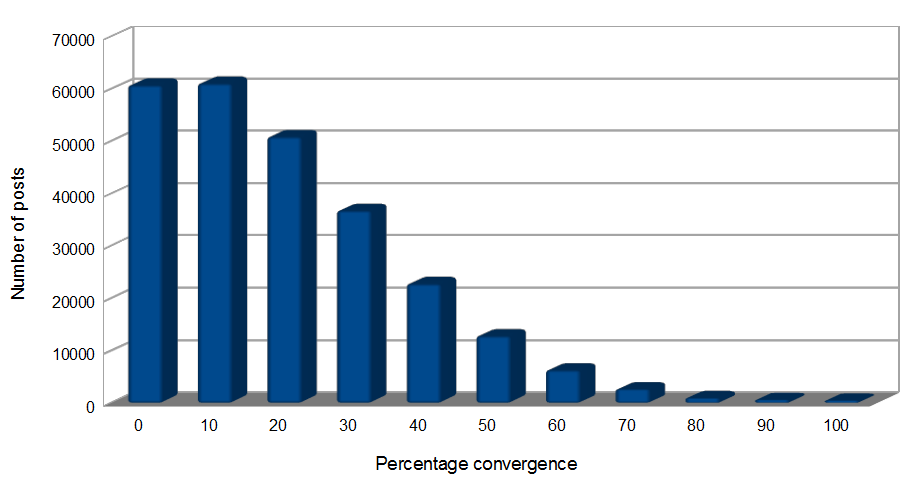}
	\caption{Convergence level of TF-IDF keywords and most significant Topic Model words for inferenced topics}
	\label{fig:tfidf-topic}
\end{figure}

Fig. \ref{fig:tfidf-topic} presents comparison between TF-IDF keywords and Topic Modelling. As it can be seen, for about 20\% of documents there was not even a single matching word. Convergence rate above 50\% was achieved by merely 6000 posts which is around 5\% of all input data.

\begin{figure}[ht]
	\centering
	\includegraphics[width=\linewidth]{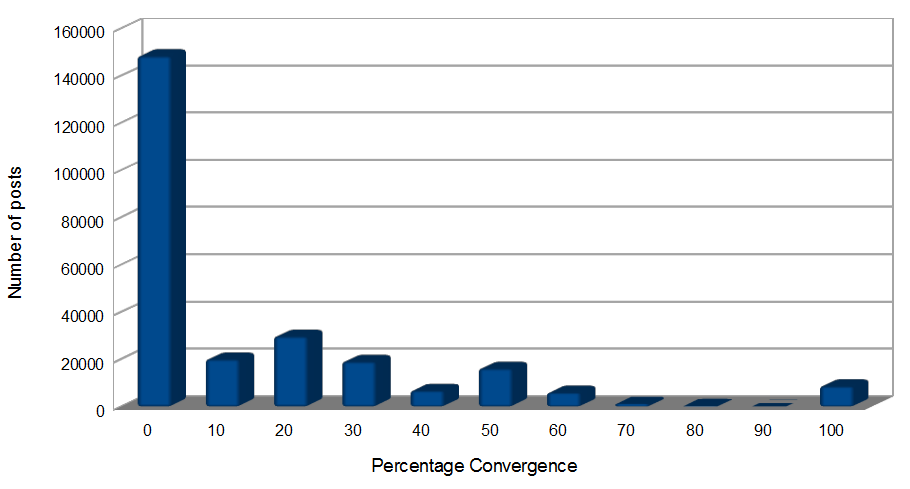}
	\caption{Convergence level of TF-IDF keywords and user provided Tags}
	\label{fig:tfidf-tag}
\end{figure}

In fig. \ref{fig:tfidf-tag} TF-IDF keywords are compared with user tags.
One can see that for huge part of documents achieved convergence rate was 0\%.
There are also local maxima at 20, 30, 50 and 100\% and their origin is connected with number of tags user provides, which in most cases is between 1 and 5 (when matched 1/5, 1/4, 1/3, 1/2 and 1/1 of keywords).

\begin{figure}[ht]
	\centering
	\includegraphics[width=\linewidth]{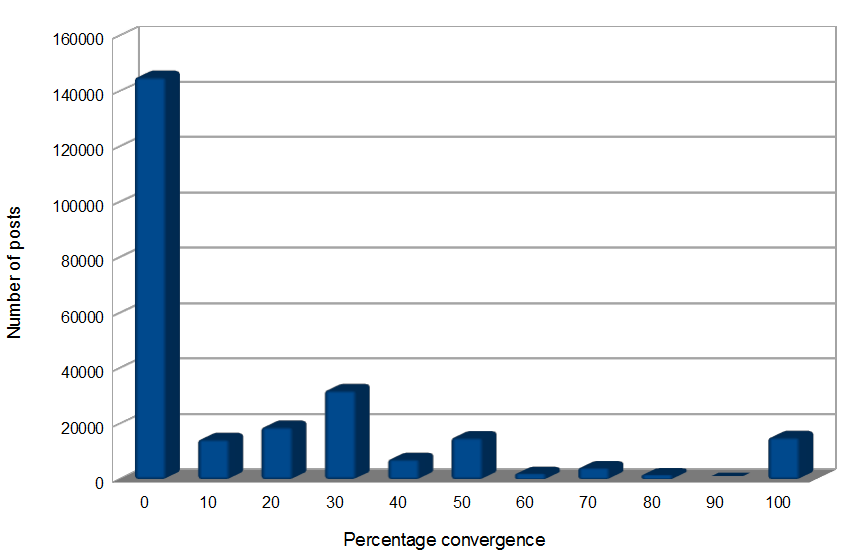}
	\caption{Convergence level of user provided Tags and most significant Topic Model words for inferenced topics}
	\label{fig:tag-topic}
\end{figure}

Fig. \ref{fig:tag-topic} displays user tags in comparison with topic modelling.
This histogram shows that, similarly to fig. \ref{fig:tfidf-tag}, very huge part of tested posts don't have even a single word in common. We can also notice similar maxima due to the same reasons.

According to presented data, convergence rate seems to be very low, and therefore it would be imperative to check which of presented methods gives correct results. However, in depth analysis of our results uncovered that in fact all three methods are properly describing posts semantic. The problem is with vocabulary that is being used to describe it. The reason for different vocabulary is connected with level of generalisation that is utilised by presented methods:

\begin{itemize}
	\item keywords provides very specific and detailed description of given document,
	\item tags provides general summary of the document and are far less specific, but error-prone of misspelling (they are provided by post authors),
	\item topics provides very general and abstract idea behind the document.

\end{itemize}

For these reasons, we are focusing more on topic modelling.


\subsection{Topics coverage by groups}

\begin{figure}[ht]
	\centering
	\includegraphics[width=\linewidth]{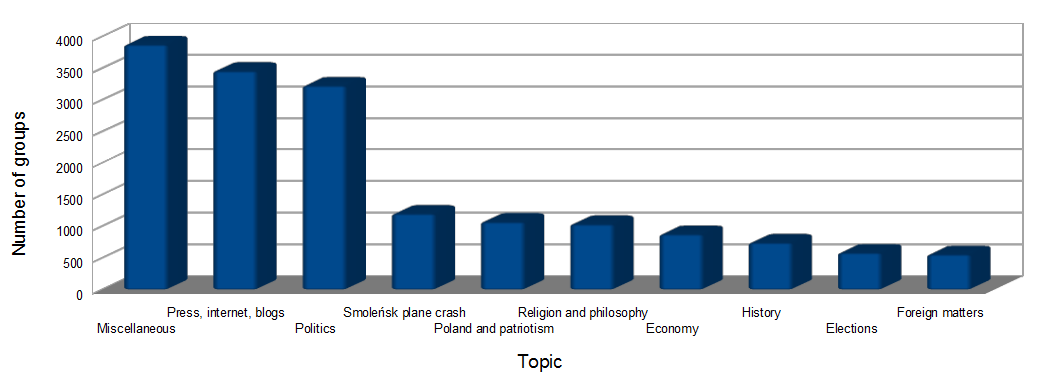}
	\caption{Number of groups exploiting given topic}
	\label{fig:topicGroupsNumber}
\end{figure}

Figure \ref{fig:topicGroupsNumber} shows number of groups that exploit given topic in more than 5\% of total group members messages. This figure presents ten most popular topics. We can notice that most popular topic is {\it Miscellaneous} which is very general and in fact is a mix of many themes. 


\subsection{Influence of group size on covered topics}
In this experiment we aimed to check if there is a correlation between groups size and topics this groups covers.

%
%
%


\begin{figure}[ht]
	\centering
	\includegraphics[width=\linewidth]{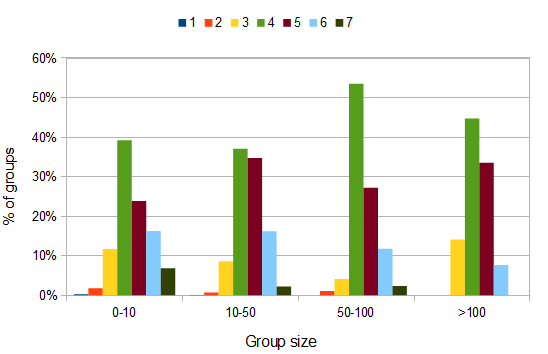}
	\caption{Number of topics covered by groups with certain size, k=5}
	\label{fig:size-topics-5}
\end{figure}

Fig. \ref{fig:size-topics-5} shows number of Topics covered by groups with certain sizes. 
Considering 5\% threshold of topic importance we used to remove noise. Maximum value any group could achieve was twenty topics, however no group covered more than seven topics and most groups cover four or five topics.
In fig. \ref{fig:size-topics-5} we can observe that three to seven topics are in groups of any size, but there are some small and medium-size groups that discuss only about one or two topics, which not happens in large groups.

\begin{figure}[ht]
	\centering
	\includegraphics[width=\linewidth]{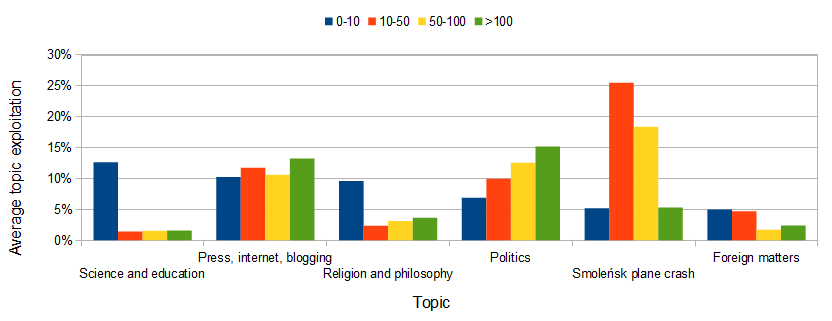}
	\caption{Percentage topics exploitation for groups with certain size, k=5}
	\label{fig:size-topics-labels-5}
\end{figure}

Fig. \ref{fig:size-topics-labels-5} presents that for some topics {\it topic exploitation} is very similar regardless of group size (e.g. {\it Press, internet, blogging} topic), but there are some specific topics that are discussed to a greater extent among members in small group (e.g. topic related with science) or among members in larger ones (e.g. politics).

\subsection{Influence of duration time on covered topics}

\begin{figure}[ht]
	\centering
	\includegraphics[width=\linewidth]{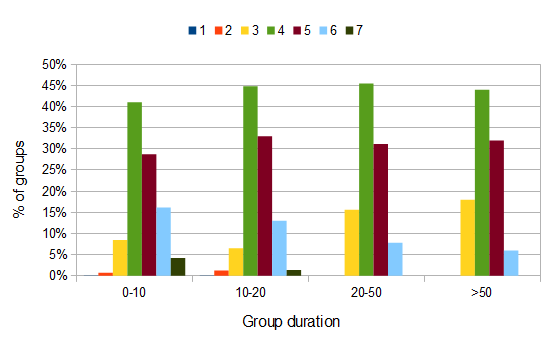}
	\caption{Number of topics covered by groups with duration time, k=5}
	\label{fig:length-topics-5}
\end{figure}

We can notice some regularities on fig.  \ref{fig:length-topics-5} :
\begin{itemize}
	\item only very short living groups covers one and two topics,
	\item as previously, most groups cover four and five topics.
\end{itemize}

\begin{figure}[ht]
	\centering
	\includegraphics[width=\linewidth]{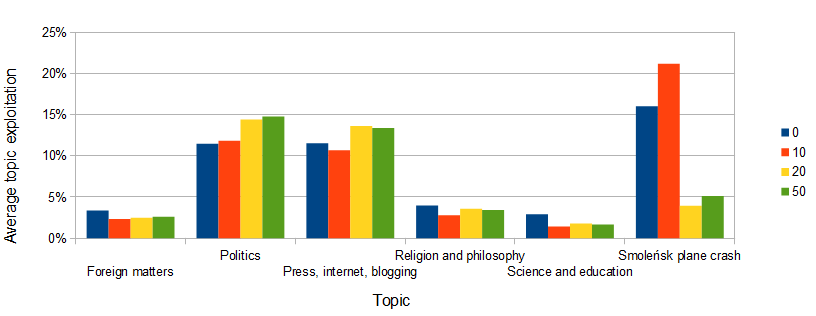}
	\caption{Percentage topics exploitation for groups with certain duration time, k=5}
	\label{fig:length-topics-labels-5}
\end{figure}

In fig. \ref{fig:length-topics-labels-5} one can see that most topics achieved similar scores, with an exception - Polish Air Force Tu-154 crash was, to a large degree, discussed by short-term groups.

%


\subsection{Connection between topics change and groups evolution}

This experiment aimed to verify if there is any apparent connection between group evolution type and change in topics coverage for this group. Experiment was based on three measures (described earlier in section \ref{topicChanges}):

\begin{itemize}
	\item average overall change in topics coverage after evolution,
	\item average maximal positive change of single topic,	
	\item average maximal negative change of single topic.		
\end{itemize}

Evolution events were taken into consideration only if there were at least ten such events in selected time period.
Presented results were collected for groups with k=5 and for periods of length 360 days.
There are two evolution types that are not present on the chart - {\it split\_merge} that did not occur and {\it decay} that was omitted. {\it Decay} event means that group ceased to exist, and therefore we cannot calculate how topics of this group changed, because there is no continuation of the group.

\begin{figure}[ht]
	\centering
	\includegraphics[width=\linewidth]{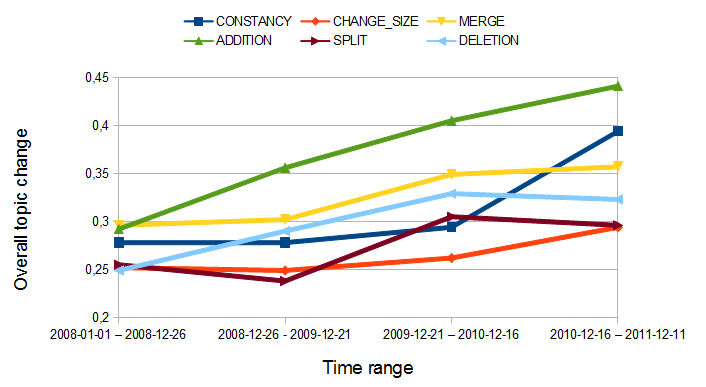}
	\caption{Average overall change in topics coverage for every evolution event type - period 360 days}
	\label{fig:avg360}
\end{figure}

{\bf Average overall change in topics}. Figure  \ref{fig:avg360} presents some regularities:
\begin{itemize}
	\item {\it addition} event has clearly the highest overall topic change,
	\item {\it split} and {\it change\_size} are connected with the lowest topic change,
	\item {\it merge} and {\it deletion} are in between,
	\item quite surprisingly {\it constancy} seems to vary between periods, even though one could expect that it will be connected with very small topic change.
\end{itemize}

\begin{figure}[ht]
	\centering
	\includegraphics[width=\linewidth]{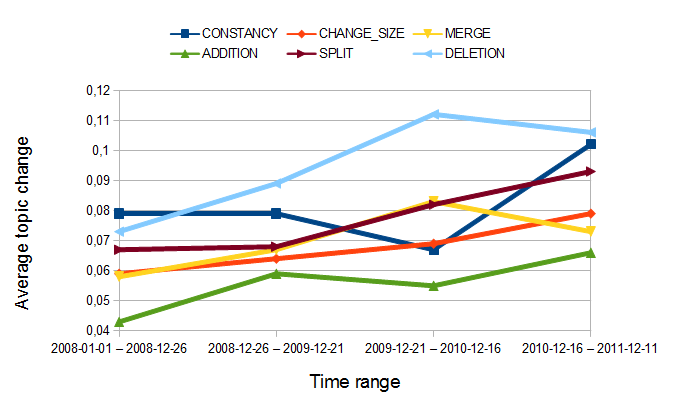}
	\caption{Average maximal positive change of single topic - period 360 days}
	\label{fig:max360}
\end{figure}

{\bf Average maximal positive change of single topic}. Figure \ref{fig:max360}  presents that:
\begin{itemize}
	\item {\it addition} has lowest average maximum single topic change. It means that on average after {\it addition} even topic that gained the most, gained very little.
	\item {\it deletion} and {\it split} caused highest positive change.
	\item {\it merge} and change\_size were in between.
\end{itemize}

\begin{figure}[ht]
	\centering
	\includegraphics[width=\linewidth]{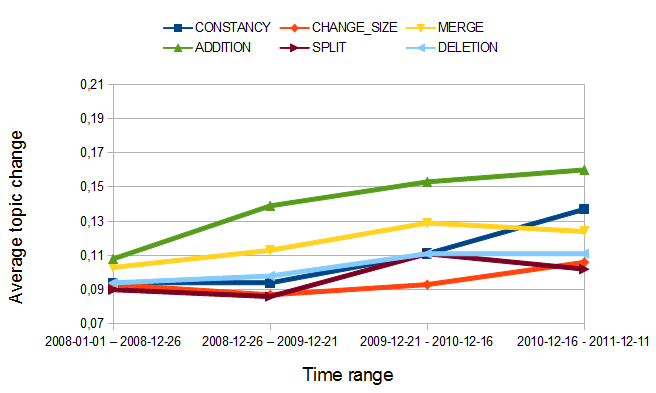}
	\caption{Average maximal negative change of single topic - period 360 days}
	\label{fig:min360}
\end{figure}

{\bf Average maximal negative change of single topic}. Figure \ref{fig:min360} shows that:
\begin{itemize}
	\item {\it addition} is connected with highest drop for a single topic. It means that after {\it addition} there is a topic that significantly loses popularity.
	\item {\it change\_size}, {\it split}, {\it deletion} and {\it merge} has very low average drop in topic popularity - even topics that lose popularity after such events lose very little.
\end{itemize}

{\bf Summary for different types of topic change}.
 {\it Addition} is connected with: highest overall change in topics, highest negative change and lowest positive change. Therefore, we can deduce that when multiple small groups are forming a single large group it is usually connected with significant drop of popularity of the main topic of each of the groups, and small rise in popularity of different topics - presumably of main topics of other groups.

{\it Deletion} and {\it split} cause small overall topic change, small negative change and large positive change. It means that splitting a group causes a rise of popularity of a single topic at the expense of all the others.

	{\it Change\_size} has very small changes in topics. We could expect that {\it constancy} behave the same way, however it does not.

{\it Merge} event causes medium rise of single topic popularity at the expense of all the others, meaning that joining groups are very similar and after join the leading topic emerges.

\subsection{Migrations between groups}

This experiment was conducted to check if migration of users between groups could be predicted using information about topic preferences of users and groups.

Important: currently analysed user does not have to be a member of the group. In fact, when calculating probability of joining a group we consider only users that are not yet members.

As a measure connected with users and groups topics we used {\it topic divergence} between user and group (details in section \ref{topicDivergence}).

Based on this measure, we tested their influence on:
\begin{itemize}
	\item probability of leaving the group (candidates for leaving group are all members of group),
	\item probability of joining the group (candidates for joining are all users that were active in the previous time slot).
\end{itemize}
%
%
%
%
%
%
%
%

While calculating joiners and leavers sets we considered all group continuations to be a single group. The reason for that is to prevent {\it deletion} event to distort results - if a group splits into multiple small groups and we are assuming that anyone from the group can leave, then we will get very high accuracy from each event when huge group changes into many small groups.
%
%

\begin{figure}[ht]
	\centering
	\includegraphics[width=\linewidth]{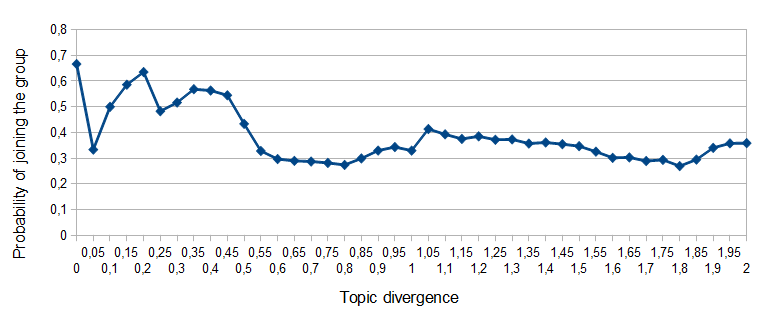}
	\caption{Probability of user joining a group based on topic divergence.}
	\label{fig:topic-join-p}
\end{figure}
\begin{figure}[ht]
\centering
\includegraphics[width=\linewidth]{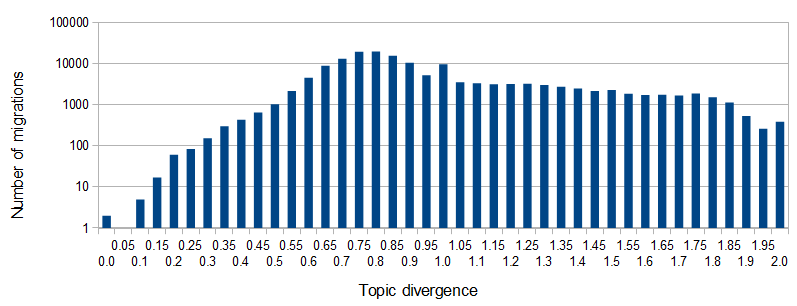}
	\caption{Number of users that did join the group based on topic divergence.}
	\label{fig:topic-join-log}
\end{figure}

{\bf Joining groups}. Figure \ref{fig:topic-join-p}  shows that there is high probability of joining for users that are high convergent with the group, however, when we look at figure \ref{fig:topic-join-log}, we can notice that some of them (most convergent users with groups) are marginal cases (very few migrations). Moreover, we can notice that probability of joining groups is rather constant regardless the value of {\it topic divergence}, except the smallest values of this measure.





Most migrations occur between 50\% and 100\% divergence and we can also see a rise in probability there, due to more cases.
There is an interesting local extreme in figure \ref{fig:topic-join-log} around 100\% divergence.
It would seem that there is a large portion of joiners that in previous time slot wrote all their posts and comments on one topic, or more likely wrote only a single post or comment, and then joined the group where this topic was not relevant.
It is worth noting that our candidates were only a fraction of real joiners. It seems that about half joiners were people that were inactive in previous time slot (151~819 joiners were inactive user and 154~977 real joiners were from the candidate set).

\begin{figure}[ht]
	\centering
	\includegraphics[width=\linewidth]{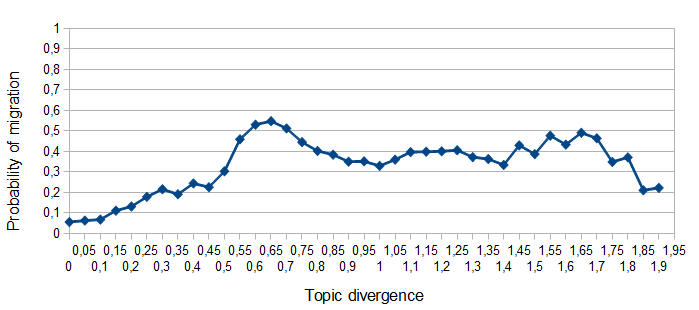}
	\caption{Probability of user leaving a group based on topic divergence.}
	\label{fig:topic-leave-p}
\end{figure}

\begin{figure}[ht]
\centering
\includegraphics[width=\linewidth]{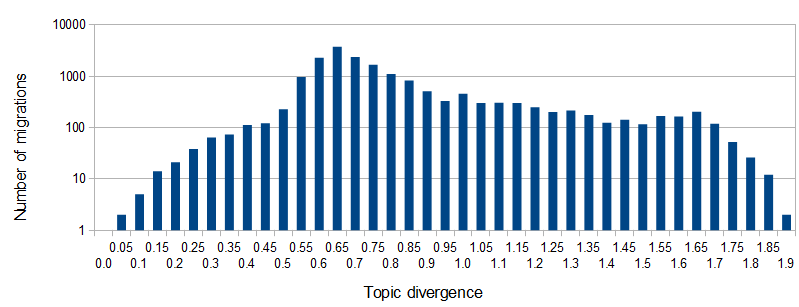}
	\caption{Number of users that did leave the group based on topic divergence.}
	\label{fig:topic-leave-log}
\end{figure}

{\bf Leaving groups}. As can be deduced from figure \ref{fig:topic-leave-p}, for low divergence values probability of leaving is low and is rising along with divergence up until 50\% divergence and, further, the probability of leaving is rather constant. Alhough, it drops down for very high divergence, however, figure  \ref{fig:topic-leave-log} tells us that it is a marginal case.





%% file: Conclusion.tex
\section{Conclusion}

In this paper the analysis of topics for communities detected in real-world data from blogosphere is presented. We conducted experiments concerning relations between discussed topics by members of groups and some aspects of groups such as their duration time or their size. Furthermore, we also analysed influence of group evolution events on changes of topics and investigated the impact of topic divergence on users behaviour such as joining or leaving group.   

Presented results seem promising and they reveal new insights into behaviour of groups and individuals. Analysis of topics discussed inside communities can be useful tool enabling better understanding of processes inside social network. 

In future we are planning to use information about topics to improve our method of prediction of group behaviour \cite{Gliwa:2013b}. Moreover, we intend to carry out similar experiments (related to analysis of topics in communities) on other datasets - we want to use also blogs in English language and datasets from different kinds of social media e.g. microblogs. Another interesting direction of further research is the analysis of key persons in groups in terms of topics they discuss and such analysis could lead to enhance our method of defining user roles and finding most influential people \cite{Gliwa:2013a}.